\documentclass[fleqn,usenatbib]{mnras}

\usepackage{newtxtext,newtxmath}

\usepackage[T1]{fontenc}

\DeclareRobustCommand{\VAN}[3]{#2}
\let\VANthebibliography\thebibliography
\def\thebibliography{\DeclareRobustCommand{\VAN}[3]{##3}\VANthebibliography}

\usepackage{graphicx}	
\usepackage{amsmath}	
\usepackage{amssymb}	
\usepackage{physics}
\usepackage{dsfont}
\usepackage{xspace}
\usepackage{hyperref}
\usepackage{orcidlink}

\newcommand{\eg}{e.g.\xspace}
\newcommand{\sn}{SN\xspace}
\newcommand{\sne}{SNe\xspace}
\newcommand{\snia}{SN~Ia\xspace}
\newcommand{\sneia}{SNe~Ia\xspace}
\newcommand{\abell}[1]{A\,#1\xspace}

\newcommand{\ie}{i.e.\xspace}
\newcommand{\manticore}{\textsc{Manticore}\xspace}
\newcommand{\skysurvey}{\textsc{skysurvey}\xspace}
\newcommand{\rateunit}{\times10^4\ \mathrm{SNe}\ \mathrm{Gpc}^{-3}\mathrm{yr}^{-1}}
\renewcommand{\arcsec}{\hbox{$^{\prime\prime}$}} 

\title[Large-Scale Structure from ZTF \sneia]{Illuminating the Local Universe: Large-Scale Structure from ZTF Type Ia Supernovae}

\author[A. Gilles Lordet et al.]{Antoine Gilles Lordet\orcidlink{0009-0002-4464-8063},$^{1}$\thanks{E-mail: antoine.gilleslordet@fysik.su.se}
Ariel Goobar\orcidlink{0000-0002-4163-4996},$^{1}$
Jens Jasche\orcidlink{0000-0002-4677-5843},$^{1}$
Stuart McAlpine\orcidlink{0000-0002-8286-7809},$^{1}$
Jesper Sollerman\orcidlink{0000-0003-1546-6615},$^{2}$\newauthor
Young-Lo Kim\orcidlink{0000-0002-1031-0796},$^{3}$
Mickael Rigault\orcidlink{0000-0002-8121-2560},$^{4}$
Madeleine Ginolin\orcidlink{0009-0004-5311-9301},$^{5}$
Umut Burgaz\orcidlink{0000-0003-0126-3999},$^{6}$
Eric C. Bellm\orcidlink{0000-0001-8018-5348},$^{7}$\newauthor
Matthew J. Graham\orcidlink{0000-0002-3168-0139},$^{8}$
Joahan Castaneda Jaimes\orcidlink{0000-0002-0987-3372},$^{9}$
Frank J. Masci\orcidlink{0000-0002-8532-9395},$^{10}$
Josiah Purdum\orcidlink{0000-0003-1227-3738},$^{8}$
Reed Riddle\orcidlink{0000-0002-0387-370X}$^{8}$
\\
$^{1}$The Oskar Klein Centre, Department of Physics, Stockholm University, Albanova University Center, SE-106 91 Stockholm, Sweden\\
$^{2}$The Oskar Klein Centre, Department of Astronomy, Stockholm University, Albanova University Center, SE-106 91 Stockholm, Sweden\\
$^{3}$Department of Astronomy \& Center for Galaxy Evolution Research, Yonsei University, Seoul 03722, Republic of Korea\\
$^{4}$Université Claude Bernard Lyon 1, CNRS, IP2I Lyon/IN2P3, IMR 5822, F-69622 Villeurbanne, France\\
$^{5}$Institute of Astronomy and Kavli Institute for Cosmology, University of Cambridge, Madingley Road, Cambridge CB3 0HA, UK\\
$^{6}$School of Physics, Trinity College Dublin, College Green, Dublin 2, Ireland\\
$^{7}$DIRAC Institute, Department of Astronomy, University of Washington, 3910 15th Avenue NE, Seattle, WA 98195, USA\\
$^{8}$California Institute of Technology, 1200 E. California Blvd, Pasadena, CA 91125, USA\\
$^{9}$Division of Physics, Mathematics and Astronomy, California Institute of Technology, 1200 E. California Blvd, Pasadena, CA 91125, USA\\
$^{10}$IPAC, California Institute of Technology, 1200 E. California Blvd, Pasadena, CA 91125, USA
}

\date{Accepted XXX. Received YYY; in original form ZZZ}

\pubyear{2025}

\begin{document}
\label{firstpage}
\pagerange{\pageref{firstpage}--\pageref{lastpage}}
\maketitle

\begin{abstract}
Within the volume-limited subsample at $z<0.06$ of the Zwicky Transient Facility (ZTF) DR2 sample, we confirm a statistically significant excess of Type~Ia supernovae (\sneia) around $z \simeq 0.02$--$0.04$, previously reported but not explained by survey selection effects. Forward simulations assuming a uniform volumetric \snia rate and realistic ZTF detection efficiencies fail to reproduce the feature, rejecting this hypothesis at a $12.2\sigma$ level for the volume-limited sample. We further detect excesses in the rates compared to our survey simulations at $z \simeq 0.08$ and $0.14$, increasing the significance to $14\sigma$ when extending the redshift range up to $0.12$. To investigate the origin of these inhomogeneities, we compare the observed \snia distribution to constrained reconstructions of the local matter density field from the \manticore project, based on Bayesian forward modelling of the 2M++ galaxy catalogue. While \sn overdensities are spatially associated with prominent nearby structures such as the Perseus, Coma, and Hercules superclusters, the amplitude of the \sn excesses significantly exceeds that expected from local matter overdensities alone. By reconstructing a redshift-dependent volumetric \snia rate, we find that local enhancements can reach factors of three to eight within specific clusters, while the sample-averaged rate remains consistent with previous low-redshift measurements. These results indicate that the \snia rate is not a linear tracer of the underlying matter density and suggest a strong environmental dependence in dense structures. We discuss possible physical origins and highlight the implications for low-redshift \sn cosmology, including correlated peculiar velocities and additional covariance beyond standard linear corrections.
\end{abstract}




\section{Introduction}

Type Ia supernovae (\sneia) are critical tools in modern cosmology, serving as standardisable candles for measuring cosmic distances and tracing the expansion history of the universe. Their uniform peak luminosities have enabled the discovery of cosmic acceleration \citep[][see \citealt{2011ARNPS..61..251G} for a review]{Riess98,Perlmutter99} and remain central to ongoing efforts to refine cosmological parameters \citep{Betoule14,Scolnic18,2022ApJ...938..110B,DES5YR,Unity3, popovic_dark_2026}. However, the observed distribution and discovery rate of \sneia may not be uniform across cosmic volumes, raising the question of whether large-scale structure (LSS) influences where and how often these supernovae are detected \citep{2022MNRAS.510..366T}.

Since \sneia originate from thermonuclear explosions of white dwarfs in binary systems which are hosted in galaxies, it is plausible that their spatial distribution and discovery rates are tied to the underlying LSS. Previous analysis have shown an increase in \snia rates in clusters compared to field galaxies \citep{mannucci_supernova_2008}. Although not all the differences are significant, this observation led to design of high-redshift survey targeting clusters to increase the statistics of observed \sne \citep{hayden_hst_2021, sand_multi-epoch_2012}.

Understanding the relationship between \snia rates and the LSS has also implications for high-precision cosmological parameters inferred from supernova samples. Several works have shown correlations between stretch and high \citep{ruppin_ztf_2025,larison_environmental_2024} or low \citep{aubert_ztf_2025} density environments, global host stellar mass \citep{ginolin_ztfstretch_2025} or galaxy morphology type \citep{senzel_ztfgal_2025}. Any rate dependency with local environment will affect how these correlations bias cosmology analysis.

In Section \ref{sec:data}, we introduce the ZTF \snia DR2 sample and the \manticore-local simulations. In Section \ref{sec:std_sims} we present simple simulations of the ZTF \snia DR2 sample under two different spatial priors, uniform and proportional to the \manticore dark matter density, and how those simulations differ from the true DR2 observations. In Section \ref{sec:rate}, we present our redshift-dependent rate estimator and compare it to previous measurement in \ref{sec:results}. Finally in Section \ref{sec:discussion}, we discuss potential origins for the observed spatial dependency of the \snia rate and its implication for low-redshift cosmology analysis.

\section{Data}
\label{sec:data}
\subsection{The Zwicky Transient Facility (ZTF)}
The Zwicky Transient Facility (ZTF) is a wide-field optical time-domain survey operating on the 48-inch Samuel Oschin Telescope at Palomar Observatory \citep{Bellm+2019, Graham+2019, Dekany+2020, Masci+2019, Patterson+2019, Mahabal+2019, Duev+2019}. Its camera covers approximately 47~deg$^{2}$ per pointing, allowing a large fraction of the northern sky to be observed with cadences ranging from nightly to a few days. This combination of large étendue and high cadence makes ZTF particularly efficient at discovering supernovae in the nearby Universe. ZTF’s wide sky coverage is also well matched to the angular extent of nearby large-scale structures. Structures like superclusters or filaments cover tens of degrees on the sky at low redshift and are therefore inefficiently sampled by narrow-field surveys. ZTF’s large footprint allows these structures to be observed coherently thus enabling the detection of spatial inhomogeneities in the \sn population associated with the local matter distribution. In addition, ZTF has a low-resolution integral field spectrograph (SEDm) dedicated to spectroscopically classifying transients detected by the photometric survey \citep{SEDm, IFU, kim_new_2022, lezmy_hypergal_2022}. The brightness of nearby events also facilitates host-galaxy identification leading to better redshift estimates from galaxy surveys such as MOSThosts survey \citep{soumagnac_most_2024}.

These properties allow ZTF to construct a large homogeneous sample of low-redshift \sneia where selection effects are negligible, making it particularly sensitive to environmental and LSS effects rather than survey-induced biases.

\subsection{ZTF \snia DR2 sample}
The ZTF \snia DR2 sample contains 3628 spectroscopically confirmed \sneia up to $z\sim0.2$ observed between March 2018 and December 2020 \citep[see][and references therein]{Rigault24}. Out of those, 2667 pass basic quality cuts on their lightcurves and are suited for cosmology analysis. The \textit{volume limited} sample is then obtained by selecting SNe at $z<0.06$ and is assumed free from non-random selection functions. Therefore features present in the volume limited sample should trace underlying properties of the global SN population.
\begin{figure}
	\includegraphics[width=\columnwidth]{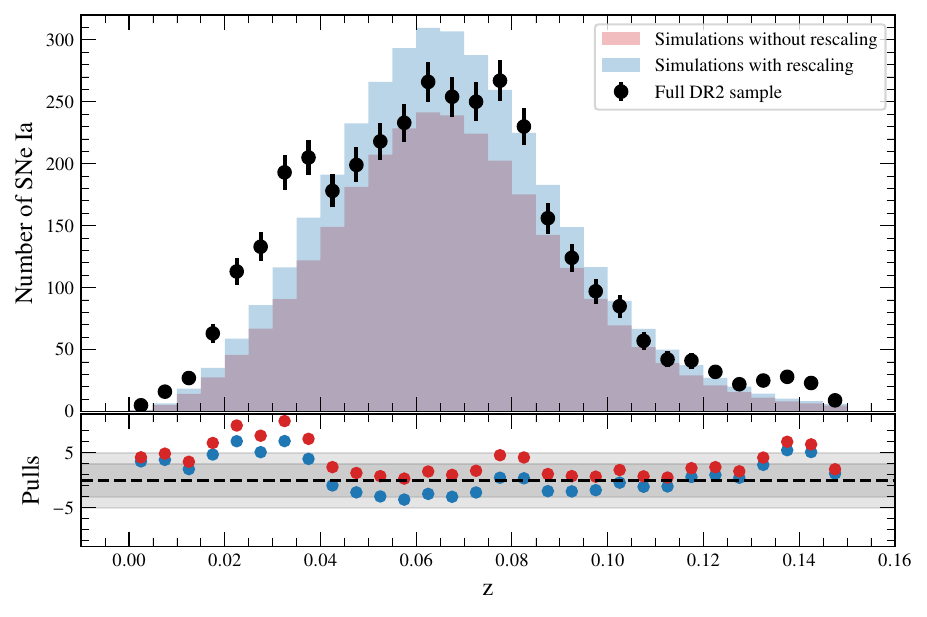}
    \caption{\textit{Upper panel}: ZTF DR2 redshift distribution for the complete \sneia sample. The shaded histograms show the average of 100 random samples for an isotropic universe and a uniform rate, either using the rate of $2.35\rateunit$ inferred by \citet{perley2020} or rescaled to match the total number of \sneia in the DR2. \\ \textit{Lower panel}: corresponding Pearson pull $(N_{obs} - N_{sim})/\sqrt{N_{sim}}$ in the redshift bins.}
    \label{fig:z_dist}
\end{figure}

The most prominent feature is a strong excess in the number of observed \sneia in the redshift range $0.02 \leq z < 0.04$ followed by a decrease for $0.04\leq z < 0.05$. This feature was previously noted by \citet{Amenouche25} and could not be accounted for using selection effects.

\subsection{Large-scale structure reconstruction in the local Universe: the \manticore project}

The observed distribution of galaxies and galaxy clusters provides an incomplete and biased tracer of the underlying matter density field. Recovering the three-dimensional large-scale structure (LSS) of the Universe therefore requires statistical reconstruction techniques that combine galaxy survey data with physical models of structure formation. In the local Universe, where non-linear gravitational evolution is important and peculiar velocities are significant, such reconstructions must go beyond linear theory to reliably infer the matter distribution on megaparsec scales.

A powerful approach for such reconstructions is Bayesian field-level inference, in which the three-dimensional initial conditions of the Universe are inferred from present-day galaxy observations \citep{jasche_physical_2019}. Rather than evolving a fixed density field forward, this framework treats the primordial density fluctuations as unknowns and samples their posterior distribution using Hamiltonian Monte Carlo (HMC). At each step of the Markov chain, a candidate realisation of the initial conditions is drawn, evolved to the present day using a fast gravitational dynamics solver (the COLA method in our case), and compared to the observed galaxy distribution through a physically motivated likelihood that accounts for non-linear galaxy bias, redshift-space distortions, survey masks, and radial selection functions. The result is a Markov chain of posterior samples over the initial conditions of the local Universe, marginalising consistently over all observational uncertainties. These posterior samples are particularly well suited for studies of nearby structure, where the signal-to-noise is high and the imprint of individual clusters, filaments, and voids can be resolved.

In this work we make use of the \manticore-local simulations \citep{mcalpine_manticore_2025}, a suite of constrained realizations of the local Universe derived from the 2M++ galaxy catalogue \citep{lavaux_2m++_2011} using the Bayesian Origin Reconstruction from Galaxies (\textsc{BORG}) algorithm \citep{jasche_physical_2019}. The 2M++ catalogue contains photometric targets from 2MASS-XSC and spectroscopic redshifts from the 2MASS Redshift Survey \citep{huchra_2mass_2012}, 6dFGS \citep{jones_6dfgs_2009} and SDSS DR7, amounting to $\sim 69\,000$ galaxies up to $z = 0.1$. After masking, the region constrained by the data spans approximately 12 to 350~Mpc in comoving distance, though the signal-to-noise ratio drops sharply beyond 200~Mpc. The constrained region thus spans most of the volume covered by the ZTF \snia sample. Each \manticore-local simulation is initialised from an independent draw from the \textsc{BORG} posterior over initial conditions, which is then evolved to $z = 0$ using a high-accuracy $N$-body solver, providing a physically self-consistent ensemble of realizations of the local dark matter distribution.

The \manticore-local simulations are generated within a cubic volume of side length 1000~Mpc and a grid resolution of $\sim 4$~Mpc, with the observer placed at the centre of the box. Prominent nearby structures such as the Virgo, Perseus, Coma, and Hercules superclusters are robustly recovered. This allows the inferred matter density field and associated halo catalogues to be used as a reference for studies of environmental effects and inhomogeneities in the ZTF \snia sample.

By providing a self-consistent probabilistic description of the local matter distribution — including its uncertainties — the \manticore-local simulations enable a direct comparison between the observed spatial distribution of \sneia and the underlying large-scale structures. This allows us to test whether the inhomogeneities observed in the ZTF DR2 \sn sample can be explained by variations in the underlying dark matter density or the presence of specific structures.

\section{Spatial distribution of Type~Ia supernovae in ZTF}
\label{sec:std_sims}
To interpret the redshift inhomogeneities observed in the ZTF DR2 \snia sample, we generate samples accounting for known selection effects under different priors for the spatial distribution of \sneia. The goal of these simulations is not to reproduce individual events but to test whether simple and physically motivated assumptions about the spatial distribution of \sneia can account for the observed features in the data. 

The simulations are constructed using the \skysurvey\footnote{https://skysurvey.readthedocs.io/en/latest/} package to match the main characteristics of the ZTF survey, namely its footprint and selection function. \sneia are generated using the distributions for their intrinsic parameters previously measured by the ZTF collaboration on the DR2 data \citep{ginolin_ztfcolor_2025, ginolin_ztfstretch_2025}. To simulate the effect of the selection function, the obtained sample is then cut using a survival sigmoid probability on the observed ZTF $r$-band peak magnitude
\begin{equation}\label{eq:sigmoid}
    S(m_\mathrm{obs}; m_\mathrm{lim}, s_\mathrm{lim}) = \frac{1}{1+e^{s_\mathrm{lim}(m_\mathrm{obs}-m_\mathrm{lim})}}
\end{equation}
For ZTF, the limit $r$-band peak magnitude $m_\mathrm{lim}^\mathrm{r}$ is $18.8$ mag and the slope of the sigmoid function $s_\mathrm{lim}^\mathrm{r}$ is $4.5$ mag$^{-1}$ \citep{Rigault24}. The obtained sample is cut to the survey footprint and rescaled to account for the effective observed area which includes the effect of the 3-days cadence cycle. This modelisation is simpler than the one used by \citet{Amenouche25} as lightcurves are not generated. Both approaches produce similar redshift distributions up to some scaling, we use our model for simplicity.

The first spatial distribution used is a simple isotropic and homogeneous prior in comoving volume, obtained assuming a uniform volumetric rate. We then introduce spatial inhomogeneities by drawing supernova positions from the halo catalogues derived from \manticore-local simulations. The following subsections describe these simulation strategies in more detail and their ability to reproduce the redshift distribution seen in the ZTF DR2 sample.

\subsection{Isotropic universe}
First we build simple simulations of the ZTF DR2 \snia sample assuming a uniform volumetric rate and rescaled to produce on average a total number of observed \sneia comparable to the 3628 of the DR2. The resulting mean distribution of 100 simulations is represented as the shaded histogram in Fig. \ref{fig:z_dist}. While the overall shape of the distribution is similar to that of the DR2 sample and the observed number of SNe matches the prediction at high redshifts, there is a clear discrepancy of $3.8$ to $7.1\sigma$ in each redshift bins within the $0.02\leq z\leq 0.04$ range. 
Comparing the observed counts per bin to simulated ones over the volume-limited regime ($z\leq0.06$, 12 bins) leads to a reduced chi-square of $\chi^2_\mathrm{d.o.f}=18.3$, equivalent to a $12\sigma$ tension using the Wilson–Hilferty transformation with $p=3$ \citep{wilson_distribution_1931}. Including all 30 bins up to $z=0.15$ leads to $\chi^2_\mathrm{d.o.f}=10.7$, equivalent to a $14\sigma$ tension. In both cases the associated p-value is vanishingly small, therefore the assumption that the redshift distribution of \sneia in the local universe is described by a uniform volumetric rate can be safely rejected. In addition the rescaling imposed is equivalent to setting the \snia rate at $2.73\rateunit$ which is abnormally high compared to other \snia rate measurements \citep{desai_atlasrate_2026, perley2020, frohmaier_real-time_2017}. This can be explained by the fact that this rescaling impacts all scales without distinction, while the observed excesses are mainly at lower redshifts. For instance simulated counts in the $0.04 \leq z \leq 0.07$ or $0.085 \leq z \leq 0.1$ ranges are in better agreement with the non-rescaled counts than the rescaled ones. Using a \snia rate of $2.35\rateunit$ \citep{perley2020} without the rescaling results in an even stronger tension with the observations, with a reduced chi-square over the full range of $\chi^2_\mathrm{d.o.f} = 20.6$ and $\chi^2_\mathrm{d.o.f} = 36.8$ within the volume complete region, respectively equivalent to $20\sigma$ and $17\sigma$. Despite these clear tensions, the non-rescaled simulations are in good agreement with the observed counts for $0.04 \leq z \leq 0.07$ and $0.9\leq z \leq 0.12$. While of smaller amplitude, regions around $z \sim 0.08$ and $z \sim 0.14$ also show some excess of \snia compared to their neighbouring bins. They are in clear tension with the non-rescaled simulations and only become consistent with simulations when considering an abnormally high volumetric average rate.

This points to either the \snia rate being intrinsically higher at low redshift or the local cosmic variance strongly impacting the observed number of \sneia. While the $z\sim 0.08$ redshift range is outside the volume complete region, it can still be constrained as the average \sn in this range has a $r$-band magnitude of the same order as $m_\mathrm{lim}^\mathrm{r}$. This is not the case for the $z\sim 0.14$ feature as it is strongly impacted by selection effects. Therefore this analysis will focus on the $z\leq 0.12$ region to avoid potential biases.

\subsection{Anisotropic universe}

If those excesses are a product of matter anisotropies in local structures, using the dark matter halos reconstructed by \manticore as a prior on drawing SN positions should allow to reproduce them. In this scenario the total number of \sne is still computed from the uniform volumetric rate of $2.35\rateunit$. The difference comes from the sampling procedure: \sne are drawn from the halo catalogues, with each halo being weighted by its mass. For simplicity we place the \sne at the centre of each halo and set their peculiar velocity to their host halo peculiar velocity, ignoring virial motion. A more detailed study of the impact of the weighting method and whether to include the virial motion is presented in Appendix \ref{app:halo_weights}. We draw 2 realizations of \sne per \manticore halo catalogue, leading to 100 drawn SN samples per sampling method.

It is worth noting that by using the halos as a proxy, matter overdensities are enhanced through two different effects. The first one is a pure scaling relation: more massive halos in \manticore are more likely to be chosen as \snia hosts. The second channel is that those massive halos show a higher concentration of smaller subhalos than the average field, leading to neighbouring positions being sampled more often.

\begin{figure*}
    \centering
    \includegraphics[width=\linewidth]{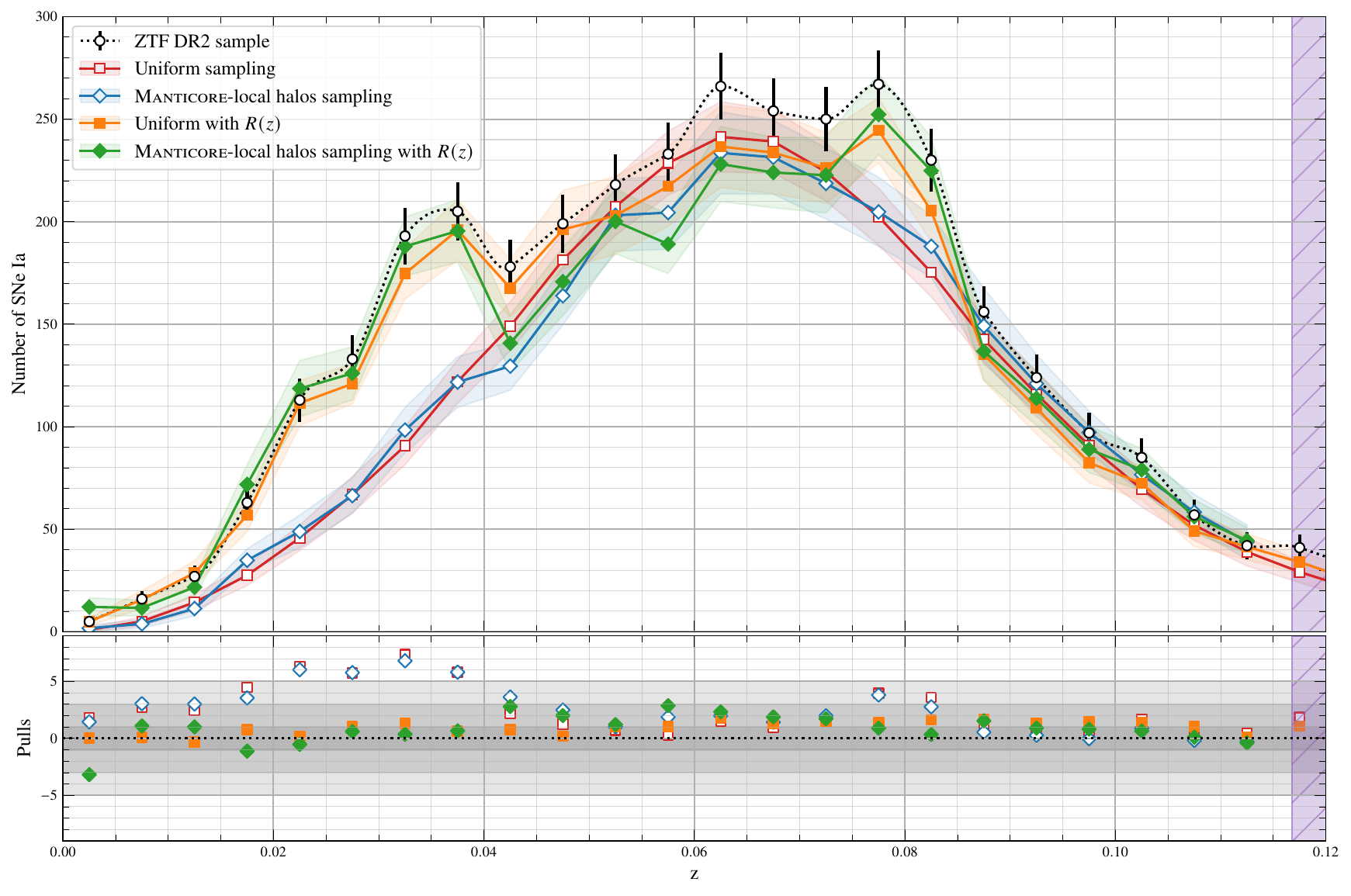}
    \caption{Simulated redshift distribution for SNe using different position priors. The shaded purple area at $z>0.11$ indicates the region not covered by the \manticore-local 1~Gpc box. Each type of simulation consists of 100 realisations. Square and diamond markers respectively represent simulations drawn i) uniformly in comoving volume and ii) from \manticore halos weighted by their masses. Additionally, unfilled markers use the comoving average rate $2.35\rateunit$ while filled markers use the redshift dependent rate inferred in Section \ref{sec:rate}.\\
    \textit{Upper panel}: The shaded regions represent the standard deviation in each redshift bin, showing the range of values taken by individual realisations. The uncertainty on the reconstructed mean per bin can be recovered by dividing the standard deviation by $\sqrt{100}$ and is of order $\sim 1$ \snia per bin.\\
    \textit{Lower panel}: Similar to Fig. \ref{fig:z_dist}, the pull per redshift bin is obtained through $(N_{obs} - N_{sim})/\sqrt{N_{obs}}$.}
    \label{fig:z_dist_sims}
\end{figure*}

The redshift distributions of the simulations based on \manticore halo catalogues are represented as blue triangles in Fig. \ref{fig:z_dist_sims}. There are fluctuations in the number of simulated events compared to the uniform sampling and the redshift range in which they occur matches the positions of the excesses observed in the DR2 sample, supporting the hypothesis that structures are the cause of those excesses. However none of those fluctuations have the required amplitude to match the observation.

\section{Redshift dependent volumetric rate and the impact of structures}
\label{sec:rate}
Quantifying the volumetric rate of Type~Ia supernovae as a function of redshift is essential for interpreting the inhomogeneities observed in the ZTF DR2 sample. While variations in the observed number of events may reflect genuine fluctuations in the underlying \snia rate, they may also arise from survey selection effects or incompleteness in spectroscopic classification and host-galaxy redshift measurements.

A previous computation of the \snia rate using ZTF data was done in \citet{perley2020}. However, their approach cannot be used here as each SN was weighted by the volume it probes with a varying limiting absolute magnitude. This does not allow for a parametrization of the rate with redshift as each volume is artificially set to the entire survey search volume. Furthermore, they used an approximation of the selection function at low magnitudes which is valid only up to 18.5 mag. Thus all events fainter than this magnitude at a given redshift would not be properly accounted for.

Another approach explored by \citet{frohmaier_real-time_2017} consists in injecting artificial SNe at the image level and using the full reconstruction pipeline to infer the survey efficiency as a function of SN parameters. While this approach naturally gives the survey efficiency as a function of redshift from which the rates can be inferred, we chose not to use it as a simpler modification of the previous model still produces consistent results.

In this work, we compute \snia rates directly from the observed ZTF DR2 sample by explicitly accounting for the effective volume probed by the survey and the probability of detecting and classifying an event at a given redshift. Rather than adopting a global rate, we allow the rate to vary with redshift to characterize the excesses of \sne seen in the DR2 by binning the \sn redshifts in bins of $\Delta z = 0.005$. This approach enables a direct comparison between the inferred rates and the expectations from uniform-rate models while allowing a more detailed description of local inhomogeneities.

In theory, to construct a redshift-dependent \snia rate, the uncertainty on the redshift needs to be small compared to the bin size. This requirement is satisfied for \sneia with an available host redshift, their redshift uncertainty being typically $\Delta z \sim 2\times10^{-5}$. \sneia in the DR2 whose redshift was determined either through galaxy emission lines using the SEDm or template matching with \texttt{snid} have a typical redshift uncertainty of $\delta z\sim0.001-0.004$ comparable to the bin size, which could in theory bias the analysis. We tested both including and not including those \sne in the reconstructed redshift-dependent rate but did not find any statistically significant difference between the two samples (see Appendix \ref{app:hostz}). We therefore decided to keep the entire sample and to not perform any cut on the host redshift availability.

While we do not perform quality cuts on lightcurves or redshift uncertainty, we restrict our analysis to $z \leq 0.12$ for two reasons. \sne at high redshift have a high magnitude while the selection function presented in Eq. \ref{eq:sigmoid} was inferred using low magnitude events \citep[see Fig. 4 in][]{Rigault24}. This becomes a problem when very faint events dominate the \sn population at a given redshift. Furthermore this cut match the volume covered by the \manticore-local simulations, as the $1$~Gpc box corresponds to a maximum redshift of $z\simeq 0.12$ and the constrained region extends up to $z = 0.1$.

We also reject \sne that are strongly affected by Milky Way dust extinction by applying a cut at $A_V > 1$~mag consistent with previous low-redshift SN rate measurements \citep{perley2020}.

A summary of the statistics of the \snia sample used is given in Table \ref{tab:cuts}.

\begin{table}
    \centering
    \begin{tabular}{cccc}
         Cuts & $\#$ SNe & Removed & \% Removed \\
         \hline
         Full sample & 3628 & -- & --\\
         $z \leq 0.12$ & 3453 & 175 & 4.8\\
         $A_V < 1$ mag & 3386 & 67 & 1.9\\
         \hline\hline
         Total & 3386 & 242 & 6.7\\
         \hline
    \end{tabular}
    \caption{Cuts applied to the ZTF DR2 \snia sample for the rate measurement.}
    \label{tab:cuts}
\end{table}

\subsection{Quantities entering the rate measurement}

The measurement of volumetric Type~Ia supernova rates relies on several observational and survey-related quantities to allow the effects of survey geometry, observation time and detection efficiency to be consistently incorporated.

In practice, the inferred rate in a given redshift interval depends on the number of detected SNe, the effective comoving volume probed by the survey in that interval, and the total effective observing time. Additional corrections are required to account for redshift-dependent detection efficiencies, incomplete spectroscopic information and time-dilation effects. The following section introduces the notation and definitions adopted throughout this work and provides the framework for the rate estimator described below.

\subsection{Rate estimate}

In this analysis we choose to model the redshift-dependent \snia rate $R_k$ in a fixed redshift range $[z_k, z_{k+1}]$ as the sum of $N_k$ events exploding within a given time span $T$ and effective comoving volume $V_{\mathrm{eff},k}$. Each event is weighted according to the likelihood of it being detected, inversely proportional to the detection efficiency $\epsilon_i$.
\begin{equation}\label{eq:rate}
    R_k = \frac{1}{V_{\mathrm{eff}, k} T} \sum_{i=1}^{N_{k}} \frac{1+z_i}{\epsilon_i}
\end{equation}
The $1+z_i$ factor accounts for cosmological time dilation between the \sn explosion and its observation.

This effective comoving volume is obtained by rescaling the comoving volume in the shell $[z_k; z_{k+1}]$ using the fraction of the sky area covered by ZTF $f_\mathrm{skyarea}$ and the reduction of this area to the low Milky-Way extinction region $f_\mathrm{ext}$. The comoving volume of the shell is obtained using the Planck 2018 cosmology.
\begin{equation}
     V_{\mathrm{eff}, k}= f_{\mathrm{skyarea}} f_\mathrm{ext} (V(z_{k+1})-V(z_k))
\end{equation}
Using the observation logs from ZTF during the period covered by the DR2 we obtain $f_\mathrm{skyarea} = 0.77$ and $f_\mathrm{ext}=0.83$.

The uncertainty on the recovered rate is obtained in a similar way using
\begin{equation}
\sigma_k = \frac{1}{V_{\mathrm{eff}, k} T}\sqrt{\sum_{i=1}^{N_k}\qty(\frac{(1+z_i)}{\epsilon_i})^2}
\label{eq:rate_err}
\end{equation}

\subsection{Detection efficiency}
\label{sec:efficiency}
There are several ways to parametrize and determine the detection efficiency of an event. Our approach aims at including a precise modelling of the \snia population while not requiring extensive computations.

We first split the detection efficiency of a given event $\epsilon_i$ into two contributions: the fraction of the total timespan the location was covered by ZTF's 3 day cadence cycle $f_{\mathrm{obstime}, i}$ and the mean detection efficiency of \sneia for ZTF at the event redshift $\varepsilon(z_i)$. The first contribution is obtained through the observation logs, by splitting them with the 3-days cadence cycle considering a location observed if it was covered at least once in any band during a cycle. The fraction $f_{\mathrm{obstime}, i}$ is then the ratio of the number of observed cycles to the total number of cycles. A more detailed explanation is given in Appendix \ref{app:cadence}.

In order to recover the mean detection efficiency of \sneia for ZTF at a given redshift, we marginalize the selection function in $r$-band presented in Eq. \ref{eq:sigmoid} over the underlying theoretical population of \sneia at a given redshift assuming the distributions of \snia intrinsic parameter (absolute magnitude, stretch, color) to be redshift independent. This choice is motivated by the relatively small volume covered by ZTF, any evolution of these distributions with redshift is negligible compared to the impact of the selection function. The stretch and color distributions are taken from previously fitted distributions on the ZTF DR2 sample \citep{ginolin_ztfcolor_2025, ginolin_ztfstretch_2025}, a bimodal Gaussian for the stretch and a Gaussian convolved with an exponential tail for the color. We denote their respective p.d.f.s $f(x_1)$ and $g(c)$. The \snia absolute magnitude at maximum in Bessell B-band is assumed to follow a Gaussian distribution centred on $M_0=-19.3$ with a standard deviation of $\sigma_\mathrm{int}=0.1$ mag accounting for the intrinsic dispersion.

For consistency with previous analysis of the ZTF DR2 sample \citep{Rigault24, Amenouche25} and in particular the inferred stretch and color distributions by \citet{ginolin_ztfcolor_2025, ginolin_ztfstretch_2025}, we used the \texttt{SALT} model \citep{guy_salt2_2007, Guy10} in its \texttt{2.4} version \citep{Betoule14} instead of the more recent model \texttt{SALT3} by \citet{Kenworthy21}.

We also include two observational effects impacting the magnitude: Milky-Way dust extinction in ZTF $r$-band and a K-correction from Bessell B~band to ZTF $r$-band. The distribution of the reduction in magnitude $A_\mathrm{r}$ due to Milky-Way dust is obtained from the $E(B-V)$ value reported in the DR2 through
\begin{equation}
    A_\mathrm{r} = 0.834 \times A_\mathrm{V} = 0.834 \times 3.1 \times E(B-V)
\end{equation}
using the conversion coefficients provided by the SVO Filter Profile Service  \citep{rodrigo_svo_2012}.
The resulting $A_r$ distribution for the volume limited sample is approximated by a shifted and scaled Lognormal distribution $h(A_r)$ (see Appendix \ref{app:ext}).

The K-correction term between the restframe Bessell B~band and observed ZTF $r$-band is computed for a given redshift $z$, stretch $x_1$ and color $c$ using the \texttt{SALT2.4} spectral surface $F(\lambda; x_1, c)$ at phase $p=0$ as
\begin{equation}
    K(z, x_1, c) = 2.5 \times \log_{10} \frac{\int T_\mathrm{Bessell\ B}(\lambda)F(\lambda; x_1, c)\dd{\lambda}}{\int T_\mathrm{ZTF\ r}(\lambda) F((1+z) \lambda; x_1, c) (1+z) \dd\lambda }
    \label{eq:kcorr}
\end{equation}

By construction of the \textsc{salt} model, the stretch and color distribution are independent and the underlying population of \sne can be represented by the joint distribution
\begin{equation}
    P(M, x_1, c, A_\mathrm{r}) = \frac{1}{\sqrt{2 \pi }\sigma_\mathrm{int}} e^{\frac{1}{2}\frac{(M - M_0)^2}{\sigma_\mathrm{int}^2}} \times f(x_1) \times g(c) \times h(A_r)
\end{equation}
The observed magnitude of a SN with a given stretch $x_1$, color $c$, redshift $z$ and line of sight Milky-way extinction $A_\mathrm{r}$ is modelled using Tripp's Formula with additional terms for the dust and K-correction
\begin{equation}
    m_\mathrm{obs}^\mathrm{r} =  M -\alpha x_1 + \beta c + \mu(z) + K(z, x_1, c) + A_r
\end{equation}
This leads to the probability to observe this SN being
\begin{equation}
    P(\mathrm{obs} | M, x_1, c, z, A_r) =  S(m_\mathrm{obs}^\mathrm{r}; s_\mathrm{lim}, m_\mathrm{lim})
\end{equation}
and the theoretical detection efficiency at a given redshift
\begin{align}\label{eq:efficiency}
\varepsilon(z) = \iiiint & P(\mathrm{obs} |M, x_1, c, z, A_\mathrm{r}) \nonumber\\ &\quad \times P(M, x_1, c, A_\mathrm{r}) \dd{M}\dd{x_1}\dd{c}\dd{A_\mathrm{r}}
\end{align}

\begin{figure}
    \centering
    \includegraphics[width=\linewidth]{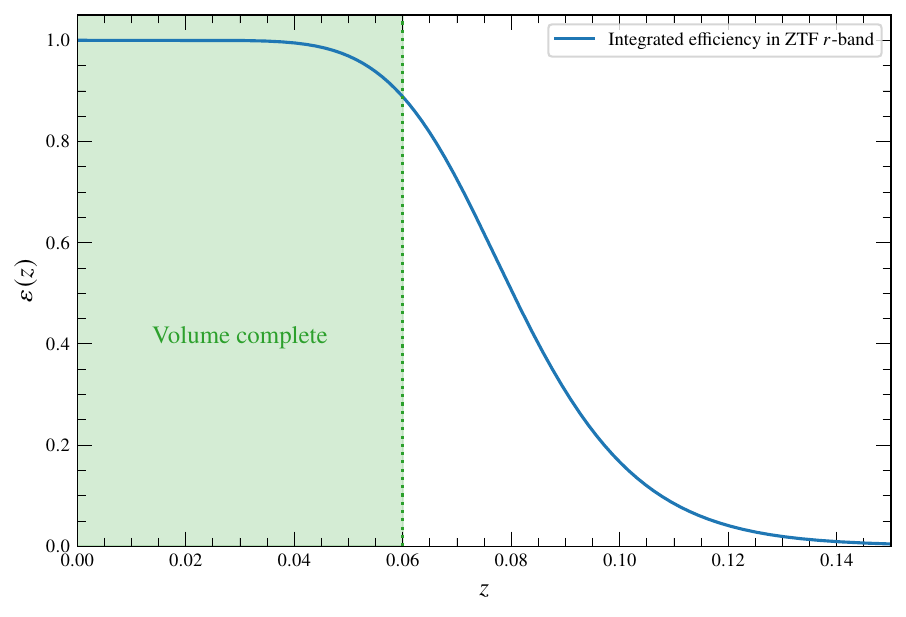}
    \caption{Theoretical mean detection efficiency of the ZTF survey for \snia events at a given redshift $\varepsilon(z)$. The volume complete region as defined in the DR2 is indicated as a reference.}
    \label{fig:ztf_eff}
\end{figure}

The obtained mean detection efficiency for ZTF is represented in Fig \ref{fig:ztf_eff}. The mean efficiency at low redshift is as expected $1.$, it drops starting at $z\sim 0.05$ and goes to zero for high redshifts. Our results indicates that the mean probability to observe a \sn at a redshift lower than $z=0.06$, not accounting for cadence variation across fields or downtime, is higher than $82\%$. This result is in agreement with the definition of the volume complete sample for ZTF, as only the most extreme events are missed and most of the normal \sneia are recovered. A more conservative cut at $z\leq0.05$ would lead to a mean probability of detection higher than $94\%$.

\section{Results}
\label{sec:results}
\begin{figure}
    \centering
    \includegraphics[width=\linewidth]{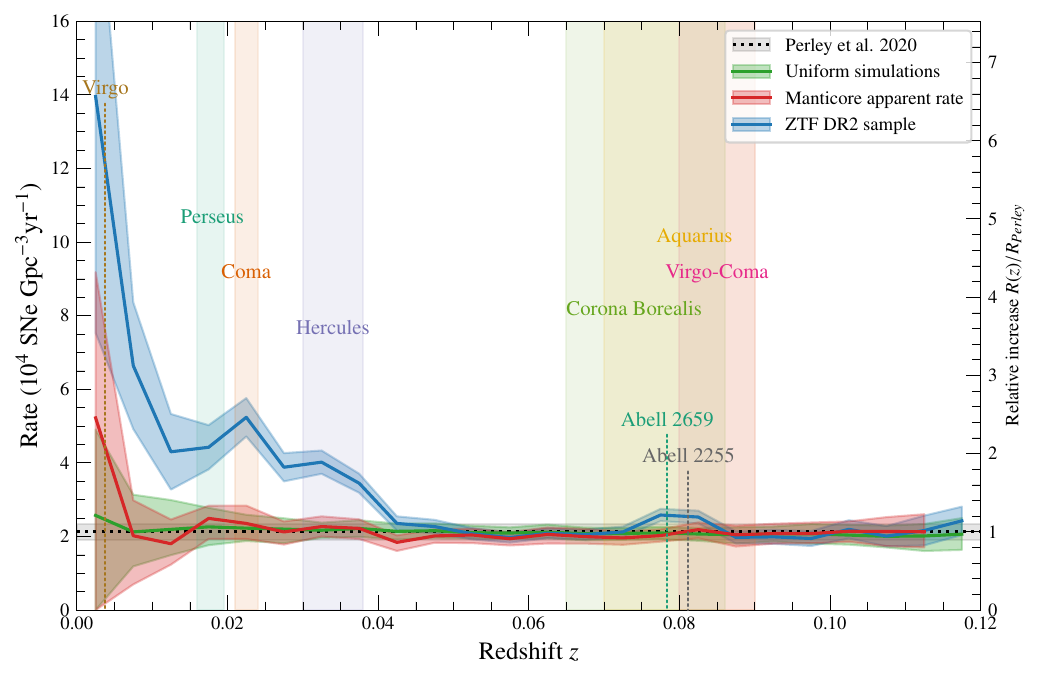}
    \caption{Redshift dependent \snia rate for (i) the DR2 sample and (ii) recovered on simulations using the uniform and \manticore-based sampling schemes. The shaded regions show the error for the DR2 sample computed using Eq. \ref{eq:rate_err} and the $0.16-0.84$ percentiles in each bin for the simulations. The average comoving rate obtained from the BTS \protect\citep{perley2020} is shown by the horizontal dashed line as a reference. Vertical shaded regions indicate the redshift range covered by known superclusters in the ZTF footprint, dashed vertical lines pinpoint more localised clusters.}
    \label{fig:dr2rates}
\end{figure}

The obtained redshift-dependent rate is presented in Fig. \ref{fig:dr2rates}. While at high redshift the rate stabilizes to a constant value around $2.2\rateunit$, the rate in the low redshift regime up to $z\sim 0.04$ is higher by a factor $1.5$ to $2$.

Using the entire $0 \leq z\leq 0.12$ redshift range as a single bin, we infer an average rate of $(2.24\pm 0.08)\rateunit$. This value is comparable to the ones inferred by \citet{perley2020} using the ZTF Bright Transient Sample (BTS) of $(2.35 \pm 0.24)\rateunit$ and \citet{frohmaier_real-time_2017} on the Palomar Transient Facility, the precursor of ZTF, at $(2.43 \pm 0.29^{+0.33}_{-0.19})\rateunit$. It is in mild tension ($2.2\sigma$) with the more recent measurement of $(2.55 \pm 0.12)\rateunit$ by \citet{desai_atlasrate_2026} on the ASAS-SN sample. The discrepancy between our inferred rate and the higher values previously inferred can be explained by the redshift dependency, as the \snia rate is high at low redshift before decreasing and stabilizing. Thus to compare the average rate we recover to earlier works we need to match the redshift range they used; for instance \citet{desai_atlasrate_2026} limit their analysis to $z\leq0.08$, for which we recover an average rate of $(2.44\pm 0.07)\rateunit$ completely resolving the previously mentioned tension.

\subsection{Re-simulation with redshift-dependent rate}

Using the inferred redshift-dependent rate, we generated a new set of simulations following the same procedure as in Sec. \ref{sec:std_sims} to verify the self-consistency of our approach.

The resulting redshift distributions, both when sampling uniformly and from \manticore halos, show a good overall agreement with the true DR2 distribution although the higher redshift bins are biased low (see Fig. \ref{fig:z_dist_sims}). As the low-redshift bins match the observation, any global scaling of the rate cannot single-handedly account for the discrepancy at higher redshifts. There are several potential causes for the slight differences between the observed and simulated distributions.

This deviation can be a consequence of using only the r~band and not considering the g~band magnitude when selecting events in the simulations. The majority of \sneia are close to the average $(c=0, x_1=0)$ and have a r~band magnitude higher than their g~band one, hence why only the ZTF r~band was used in the simulations and analysis. However this can lead to an underestimation of the detection efficiency and the observed number of \sneia for bluer events. This effect is more pronounced for events at redshift $z\gtrsim 0.06$ as the peak of the emission in the observer frame overlaps with the ZTF g~band filter.

Furthermore the mean active sky fraction we recover $f_\mathrm{eff\ area} = 0.38$ is slightly higher than the value $0.35$ previously used in \citet{perley2020} (see Appendix \ref{app:cadence}). This change in mean active sky fraction could lead to a $\sim10\%$ effect on the simulated count.

A more important effect neglected until now is the angular dependency of the rate itself. The redshift dependency introduced is difficult to interpret in term of physical effects. The high rate at low redshift $z<0.04$ would translate into a global increase of the rate over the entire sky, not matching current models tying \snia rates to star-formation history at late times. If, however, specific structures are the cause of the observed \sn excesses, we can expect specific angular locations to show high local rates while the majority of the sky does not significantly deviate from the average. Populating a redshift shell using the mean rate will enhance the number of \sne uniformly instead of tying it to those structures, thus increasing the number of \sne that are rejected by the angular cut $A_\mathrm{V} < 1$ mag and lowering the average detection rate due to more \sneia being obscured by Milky-Way dusts, especially for $z \geq 0.06$. This effect can be further affected by the previously mentioned change in mean active sky fraction depending on the coverage and mean cadence of fields with a high \sneia rate.

\subsection{Clusters and local rates}

\begin{figure*}
    \includegraphics[width=2\columnwidth]{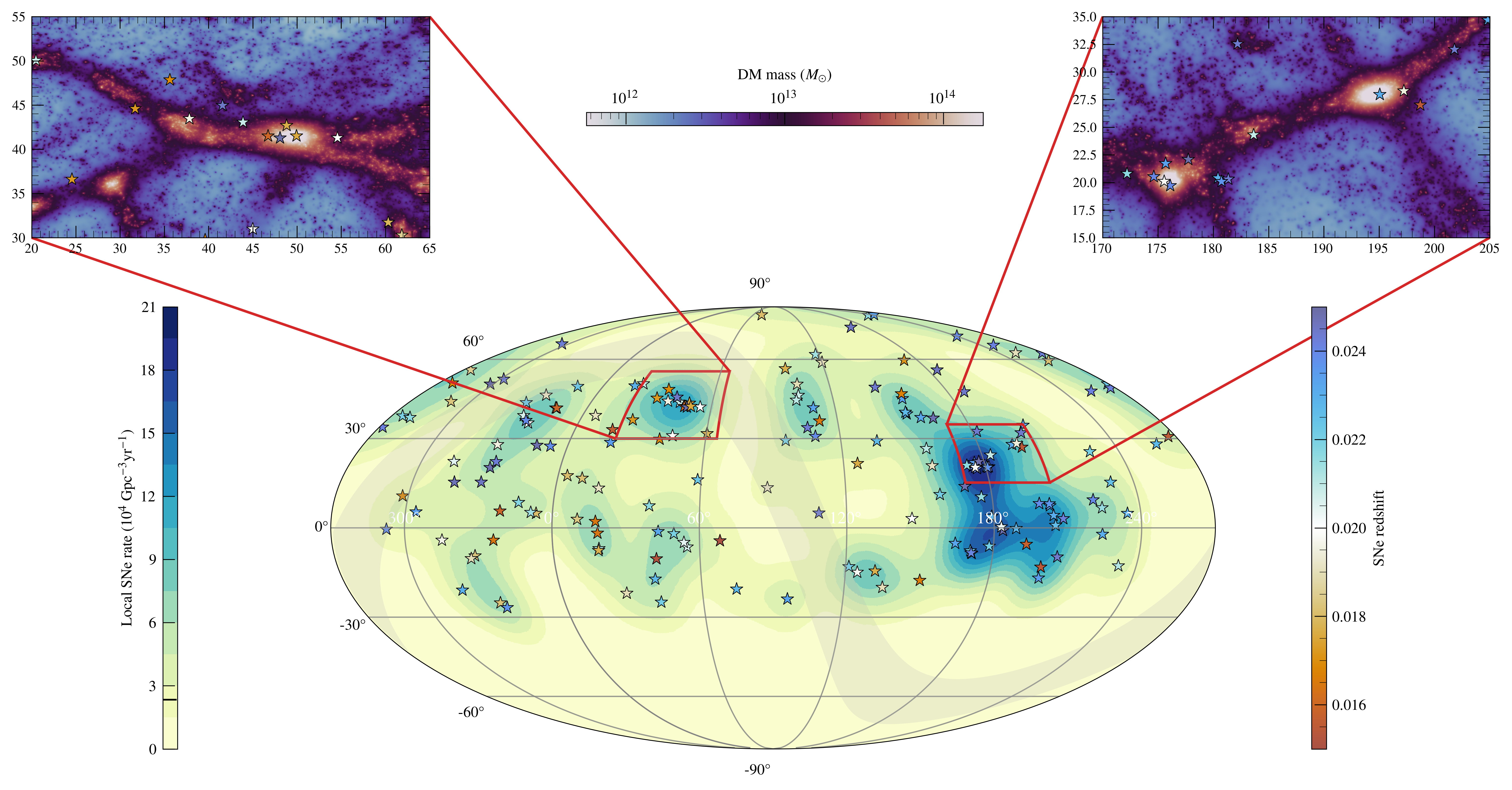}
    \caption{\textit{Skymap}: ZTF DR2 events and corresponding local rates in the redshift slice $0.015\leq z < 0.025$. Individual \snia are colored by their redshift. \textit{Upper left}: DM distribution inferred by \manticore for the Perseus Supercluster. \textit{Upper right}: DM distribution inferred by \manticore for the Coma Supercluster. ZTF DR2 \sneia are strongly clustered around the Leo Cluster (\abell{1367}) and the Coma Cluster (\abell{1656}).}
    \label{fig:0.015-0.025 slice}
\end{figure*}

\begin{figure*}
    \includegraphics[width=2\columnwidth]{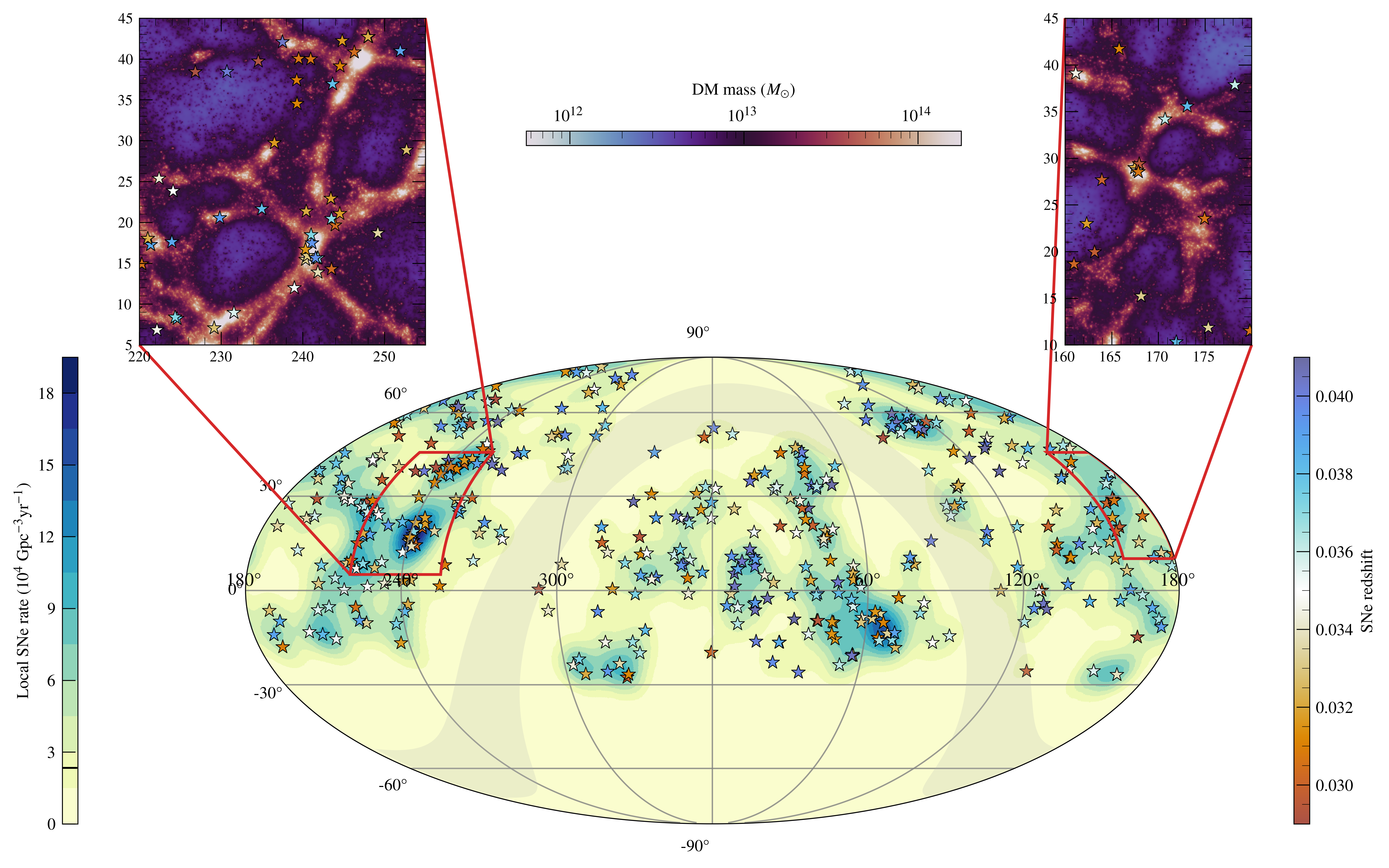}
    \caption{\textit{Skymap}: ZTF DR2 events and corresponding local rates in the redshift slice $0.03\leq z < 0.04$. Individual \snia are colored by their redshift. \textit{Upper left}: DM distribution inferred by \manticore for the Hercules Supercluster. ZTF DR2 \sneia are clumped around Hercules a (\abell{2147}, \abell{2151}). \textit{Upper right}: DM distribution inferred by \manticore for the Leo Supercluster. There are less SNe in this field, despite only 3 matching the main cluster \abell{1185} it still shows a higher than average rate.}
    \label{fig:0.03-0.04 slice}
\end{figure*}

A deeper look at the distribution of the rate in spherical shells reveals that the observed \sn excesses are not the product of an overall scaling of the rate but rather some well-localised overdensities dominating the number of observed SNe (see Figs. \ref{fig:0.015-0.025 slice} and \ref{fig:0.03-0.04 slice}).

Given that the cosmology analysis of the DR2 sample is still underway, we cannot rely on precise distance measurements to assign galaxies to superclusters and clusters or their halo counterpart in \manticore. Instead we rely only on angular coordinates within fixed redshift shells, setting them to encompass the range covered by known superclusters.

To obtain a map of the local rate in a given redshift shell, we first select the SN that belongs to this shell and assign them to the \textsc{HEALPix}\footnote{\href{http://healpix.sf.net}{http://healpix.sf.net}} pixel they land in. We then perform the per-pixel local rate using the same efficiencies (Eq. \ref{eq:efficiency}) and rate (Eq. \ref{eq:rate}) as before, only replacing the effective volume with the volume within a pixel in the given redshift shell making use of the equal-area \textsc{HEALPix} pixelisation scheme \citep{2005ApJ...622..759G}
\begin{equation}
    V_\mathrm{pixel} = \frac{V(z_{k+1})-V(z_k)}{N_\mathrm{pixel}}
\end{equation}
This volume does not correct for the fraction of the sky covered by ZTF or the reduction of the effective area due to extinction as these corrections are an overall scaling and should not be compensated locally on the sky. We then smooth the obtained map with a Gaussian kernel of effective physical size $30$~Mpc to account for the data scarcity and get an average over whole superclusters.

While cross-shell contamination is negligible for the redshift-dependent rate measurement as stated in Appendix \ref{app:hostz}, it can become an issue for the low statistics involved in specific sky regions of a given shell. To correct for potential contaminations due to poorly measured redshifts, we resample them by randomly displacing the observed redshifts according to their uncertainties assuming their distributions to be Gaussians. The skymaps shown in this analysis are the mean for $N=500$ of such resampled catalogues.

The most prominent shells for the very local universe are the shells at $0.015\leq z \leq 0.025$ and $0.03\leq z \leq 0.04$, presented in Figs. \ref{fig:0.015-0.025 slice} and \ref{fig:0.03-0.04 slice}. The insets show the  local dark matter density reconstructed by \manticore around four major structures: the Coma, Perseus, Hercules and Leo superclusters. We defined the shells for these structures using the redshifts assigned to their Abell clusters in the MSCC catalogue \citep{chow-martinez_two_2014}. The rate reported in this section are obtained from the smoothed rates skymaps in the corresponding shell and not from an assignment of \sneia to clusters. Therefore it should be interpreted as an illustration of the variance of volumetric rates depending on the environment rather than exact measurements of rates associated to specific structures.

In the MSCC catalogue, the Coma Supercluster is mainly composed of the Leo Cluster (\abell{1367}) and Coma Cluster (\abell{1656}), with the addition of three smaller clusters (\abell{1100A}, \abell{1177A}, \abell{1179A}). In term of \snia rates, the Coma cluster show a local rate of $(10 \pm 3.2) \rateunit$ while the Leo Cluster reaches $(19.7 \pm 5.1)\rateunit$.

At a similar redshift range, the Perseus Supercluster composed of the Perseus Cluster (\abell{426A}), \abell{262} and \abell{347} shows a local rate of $(10.3\pm 2.9) \rateunit$.

The Hercules Supercluster can be split into three substructures a, b and c. Hercules a (\abell{2151}, \abell{2147}, \abell{2152A}, \abell{2153A}, \abell{2159A}) dominates the entire redshift slice with a local \snia rate of $(19.0 \pm 4.5) \rateunit$. While less pronounced the local rate of \sneia is also peaking around Hercules b (\abell{2197}, \abell{2199}, \abell{2162A}) at $(9.7 \pm 2.7) \rateunit$, and Hercules c (\abell{2063A}, \abell{2052}, \abell{2040A}, \abell{2033A}) $(8.0 \pm 2.7) \rateunit$.

While occupying a similar angular surface and redshift range as the whole Hercules Supercluster, the Leo Supercluster shows a lower local rate at $(8.4 \pm 3.1) \rateunit$ with \sneia clustering mostly around the main member \abell{1185A} and some spread in the neighbouring clusters (\abell{1177B}, \abell{1179B}, \abell{1228A}, \abell{1257A}, \abell{1267A}). The Leo Supercluster provides a great counterpart to Hercules to study how the local environment affects \snia rates and population properties.

Some high local rate regions coincide not with superclusters but isolated Abell clusters, \eg \abell{496} and \abell{576} visible in Fig. \ref{fig:0.03-0.04 slice}. Respectively at $(4^\mathrm{h}33^\mathrm{m}37\fs8, -13\degr15\arcmin43\arcsec)$ and $(7^\mathrm{h}21^\mathrm{m}23\fs28, +55\degr44\arcmin38\farcs4)$ they have local \snia rates of  $(14.6\pm 4.5) \rateunit$ and $(11.3 \pm 3.4) \rateunit$. In the more distant slice $0.075\leq z \leq 0.08$, \abell{2255} has a local rate of $(12.8\pm 4.8)\rateunit$.

\section{Discussion}
\label{sec:discussion}

The ZTF \snia DR2 sample reveals statistically significant inhomogeneities in its redshift distribution, most notably the excess at $z \sim 0.03$--$0.04$. These features persist across different subsamples and quality cuts, and cannot be reproduced by uniform-rate simulations incorporating realistic survey selection effects. The central finding of this analysis is that under the assumption of a comoving average \snia rate, matter overdensities in the relevant redshift shells can typically cause an rate increase of only $10$--$20\%$, while the observed SN rate excesses in the same volumes in enhanced by a factor at least two, and up to several within well-localised individual clusters or superclusters. The \snia rate is not a linear tracer of the underlying matter density. In this section, we discuss the observational and physical origins of these inhomogeneities, present independent evidence for their non-cosmological origin, and argue that the implications for low-redshift SN cosmology are significant and demand further investigation.

\subsection{Survey-related effects}

Several observational systematics could, in principle, give rise to apparent redshift-dependent inhomogeneities, and we consider these first before proceeding to physical interpretations. The non-uniform cadence of ZTF is an obvious candidate, in particular considering the large angular extent of nearby structures. Fields towards $\alpha \sim 200^\circ$, $\delta \sim 30^\circ$ have been observed more than those at low declination or towards $\alpha \sim 160^\circ$. For low-redshift \sneia, which are typically well above the detection threshold, cadence variations will primarily affect the detection efficiency through the effective observation duration of the region on the sky, and are straightforward to account for (see Appendix \ref{app:cadence}). The cadence variations seen in the ZTF logs can account for at most a $\sim 1.6$ time increase in local rate, far from the $\sim 5$ times observed one. Reproducing the excess at $z \sim 0.03$--$0.04$ through cadence-related effects alone would thus require unrealistically strong and localised enhancements, which would necessarily imprint additional features at other redshifts or sky locations that are not observed.

Similarly, incompleteness in spectroscopic classification or host-galaxy redshift availability cannot account for the effect. \citet{Amenouche25} previously noted that they could not reproduce the inhomogeneities through selection effects alone. As discussed in Sec.~\ref{sec:rate}, including or excluding \sne without available host redshifts does not significantly change the resulting rates. Having ruled out instrumental and selection-related origins, we now turn to the physical interpretation.

\subsection{Connection to large-scale structure}

A natural interpretation of the observed excesses is that they are amplified counterparts to variations in the local matter distribution. The angular clustering of supernovae in the overdense redshift slices shows clear associations with known nearby structures, including, but not limited to, the Perseus, Coma and Hercules superclusters as well as individual Abell clusters. This observation supports a link between the \snia distribution and large-scale structure.

However, a quantitative comparison with the matter density field reconstructed by the \manticore-local simulations reveals a substantial mismatch in amplitude (see Fig. \ref{fig:dr2rates}). Assuming the \snia rate can be approximated on large scales by a constant comoving rate and that structures only affect the actual distribution of events, the effect of matter overdensities on the rate only reach the $10$--$20\%$ level relative to the cosmic mean -- excluding the lowest redshift bin dominated by Virgo where the increase can reach a factor two. In the same volume, the observed \sn excesses in shells correspond to rate increases of more than $100\%$ and up to factors of three to eight within specific clusters. Moreover, structures of comparable mass and size exhibit markedly different \snia efficiencies: some produce a large number of SNe, like the Hercules supercluster or \abell{496}, while others remain consistent with the field rate. In the Coma supercluster the less massive Leo Cluster (\abell{1367}) exhibits a higher rate than the more massive Coma Cluster (\abell{1565}), which cannot be reproduced using \manticore (see Appendix \ref{app:angular_manticore}). These results establish that a simple scaling of the \snia rate with local matter density or halo mass is insufficient to explain the observations. The excess production of \sneia in dense environments must have a physical origin rooted in the stellar populations and star-formation histories of those structures.

\subsection{Environmental dependence of the \snia rate}

A more plausible explanation is that the \snia rate is enhanced in specific environments through differences in the galaxy populations they contain.

Previous studies have already shown that the \snia rate is related linearly to both the host masses and star formation rates \citep{mannucci_supernova_2005, scannapieco_type_2005, Sullivan06}. These observations have led to the "A+B" model, where the A component scales with the host mass while the B component scales with the host star formation rate. These two terms are physically motivated by the difference in delay time of the two main channels for the \snia progenitors, the "prompt" delay time component coming from the star formation rate and the "delayed" component coming from the stellar mass. A more recent analysis by \citet{smith_sdss-ii_2012} of the SDSS-II SN Survey has further shown a preference for a bivariate power law over a linear scaling of the two components. In this framework it is expected that using only the dark matter density cannot fully reproduce the volumetric \snia rate. The observed increase in specific structures at low redshift would point to either a high star formation rate in these structures or the need to further refine the "A+B" model to include additional host properties.

Despite occupying similar volumes, the Leo and Hercules superclusters differ substantially in galaxy number counts and population mix \citep{kopylova_analysis_2011, kopylova_investigation_2013}, which may translate directly into different SN production efficiencies. In addition, highly dynamic environments such as Abell 2255 -- a cluster known for its high radio emission \citep{botteon_beautiful_2020} and elevated star-formation and AGN activity due to a recent merger \citep{miller_abell_2003} -- may further boost the SN rate through merger-triggered star-formation.

Taken together, these observations point towards an environmental modulation of the delay-time distribution (DTD) or the SN progenitor efficiency: the fraction of stellar mass converted to \sneia is not universal but depends on local conditions including metallicity, dynamical state, and binary fraction.

\subsection{Independent evidence for non-universal \snia efficiency}

The environmental dependence suggested by our observations finds independent support from an entirely different method and environment. Recent abundance-based modelling of Local Group dwarf spheroidal galaxies by \citet{2026arXiv260222333H} provides strong evidence that the Type~Ia supernova efficiency is not universal. Using chemical evolution constraints from the observed star-formation histories and metallicity distribution functions of Sculptor and Fornax, they infer a specific \snia rate per unit stellar mass approximately five times higher than the canonical field value derived from transient surveys, alongside a steep delay-time distribution, $\Psi(t) \propto t^{-2}$.

The significance of this result for our work is direct: their inferred \snia efficiency is comparable to values inferred from the \snia overdensities in the high-density LSS regions identified in this paper. The two analyses approach the same question from opposite ends of the mass and density spectrum using completely independent methodologies: low-mass, metal-poor dwarf spheroidals versus massive superclusters. Together, these independent lines of evidence substantially strengthen the case that the commonly adopted field DTD does not capture the full diversity of \snia production across environments, with consequences for iron enrichment histories and the redshift evolution of the volumetric \snia rate. Further studies including specific environment properties like stellar age, metallicity, binary fraction, or dynamical conditions are required to understand and explain the \snia rate dependency with environment.

\subsection{Implications for low-redshift SN cosmology}

The results presented here have direct and significant implications for low-redshift SN cosmology that we believe warrant urgent further investigation.

\sne originating from the same structures share correlated peculiar velocities, since superclusters are not at rest with respect to the Hubble flow. Peculiar velocity dispersions in supercluster environments have been reported to exceed those of isolated clusters or field galaxies \citep{kopylova_peculiar_2014}, extending beyond the linear peculiar-velocity corrections commonly applied in cosmological analyses.

More critically, the clustering of \sneia around massive structures means that low-redshift SN samples used to calibrate the distance ladder are not drawn uniformly from the Hubble flow. Galaxies selected as Cepheid host calibrators on the basis of having hosted bright, nearby \sneia are preferentially drawn from exactly the overdense environments identified in this work — environments where the SN rate is enhanced by factors of three to eight relative to the cosmic mean. If the peculiar velocity covariance in those environments is not fully captured by current correction schemes, or if the effective selection function of the calibrator sample is modelled as spatially uniform when the true SN rate is not, measurements of the Hubble constant from the local distance ladder may carry a systematic contribution from environmental selection bias. This is not a speculative concern: the comparison between SN-free and SN-based distance ladder measurements already shows a $\sim 1.8\ \mathrm{km}\, \mathrm{s}^{-1}\,\mathrm{Mpc}^{-1}$ offset in the inferred $H_0$ \citep[e.g.][]{stiskalek_18_2026}.

Several works have further shown correlations between host galaxy properties and \snia intrinsic parameters \citep{ginolin_ztfstretch_2025, burgaz_ztf_2026, senzel_ztfgal_2025}, although the physics at play remains unclear. A higher production of \sne in specific environments could both bias the distributions of intrinsic parameters and introduce additional correlations between events from shared environments.

If unaccounted for, such effects could bias measurements of the local expansion rate and lead to underestimated uncertainties through implicit assumptions of statistical independence between supernovae drawn from the same structure. Quantifying this environmental contribution to the systematic budget of low-redshift SN cosmology requires joint modelling of the SN rate, the large-scale density field, and the peculiar velocity covariance, exactly the combination that the \manticore framework and analyses such as this one make tractable for the first time. The present result should be understood as a first detection of the problem, not its resolution.

\section{Conclusions}
\label{sec:conclusions}
In this work, we investigated the redshift inhomogeneities of the ZTF \snia DR2 sample. Using \skysurvey, we simulated realizations of the DR2 sample under different spatial priors.
Our main findings are the following:
\begin{enumerate}
    \item Assuming a constant average rate of \sneia, we could not reproduce the spatial features of the DR2 sample, whether using an homogenous isotropic sampling or the halo catalogues from the \manticore-local simulations including information about structures in the local universe. The role played by star formation rate in the spatial prior of \snia is currently under investigation and will be the focus of a follow-up paper.
    \item We tested whether a redshift-dependent rate could be used to account for the inhomogeneities in the DR2 sample. This showed that the very nearby universe is more efficient at producing \sne than the cosmological average, and that this higher production is tightly tied to the presence of structures and clusters. However, this approach did not allow us to produce simulations fully matching the spatial distribution of the DR2 sample. Our redshift-dependent rate can only reproduce the redshift distribution, combining it with the \manticore-local halo catalogues allows to partially recover the angular distribution but not the correct 3D distribution.
    \item Excesses in the redshift distribution of \sneia should not be thought of as redshift-dependent features but as the imprint of local overdensities and structures. There is no global increase in the rate of \sneia at low redshift, rather tightly bound structures dominating the \snia rate while fields and void are less active. As such the excesses seen in the DR2 are the products of local overdense structures further enhanced by a highly non-linear relation between matter density and \sne rate.
    \item These localised and clustered excesses of \sne can have an impact on cosmological analysis through added correlations. These correlations include coherent peculiar velocities due to the structures not being at rest with respect to the Hubble flow, or enhancement of specific populations of \sneia through a shared common environment. This highlights the need for further investigations of correlations in \snia samples and how they affect cosmological analysis.
    \item This clustering of \sneia makes them a biased tracer of the LSS, therefore the relationship between \sne and the local mass needs to be further studied before using them for reconstruction of the LSS.
\end{enumerate}
We emphasize that the present analysis does not quantify any resulting bias on the Hubble constant. Evaluating such an effect would require a dedicated forward-modelling approach incorporating the environment-dependent \snia rate, survey selection functions, and the local velocity field.
Our results should therefore be interpreted as identifying a potential systematic that warrants further investigation.

\section*{Acknowledgements}
Based on observations obtained with the Samuel Oschin Telescope 48-inch and the 60-inch Telescope at the Palomar Observatory as part of the Zwicky Transient Facility project. ZTF is supported by the National Science Foundation under Grants No. AST-1440341 and AST-2034437 and a collaboration including current partners Caltech, IPAC, the Weizmann Institute of Science, the Oskar Klein Center at Stockholm University, the University of Maryland, Deutsches Elektronen-Synchrotron and Humboldt University, the TANGO Consortium of Taiwan, the University of Wisconsin at Milwaukee, Trinity College Dublin, Lawrence Livermore National Laboratories, IN2P3, University of Warwick, Ruhr University Bochum, Northwestern University and former partners the University of Washington, Los Alamos National Laboratories, and Lawrence Berkeley National Laboratories. Operations are conducted by COO, IPAC, and UW.
SED Machine is based upon work supported by the National Science Foundation under Grant No. 1106171. The ZTF forced-photometry service was funded under the Heising-Simons Foundation grant \#12540303 (PI: Graham). This work has been supported by the research project grant “Understanding the Dynamic Universe” funded by the Knut and Alice Wallenberg Foundation under Dnr KAW 2018.0067, {\em Vetenskapsr\aa det}, the Swedish Research Council, project 2020-03444.
Y.-L.K. was supported by the Lee Wonchul Fellowship, funded through the BK21 Fostering Outstanding Universities for Research (FOUR) Program (grant No. 4120200513819) and the National Research Foundation of Korea to the Center for Galaxy Evolution Research (RS-2022-NR070872, RS-2022-NR070525).
MG is supported by the European Union’s Horizon 2020 research and innovation programme under European Research Council Grant Agreement No 101002652 (BayeSN; PI K. Mande)
U.B. is funded by Horizon Europe ERC grant no. 101125877.
SM acknowledges support from the Simons Foundation through the Simons Collaboration on "Learning the Universe".
Parts of the results in this work make use of the colormaps in the CMasher package \citep{2020JOSS....5.2004V}. The results in this paper have been derived using the \texttt{astropy} \citep{astropy_collaboration_astropy_2022}, \texttt{healpy} \citep{zonca_healpy_2019, 2005ApJ...622..759G}, \texttt{dustmaps} \citep{green_dustmaps_2018}, and \texttt{scipy} \citep{virtanen_scipy_2020} packages.

\section*{Data Availability}

The ZTF DR2 \snia sample can be found at \href{https://ztfcosmo.in2p3.fr}{https://ztfcosmo.in2p3.fr/}. Informations about the Manticore simulations and data access can be found on the Aquila Consortium page \href{https://aquila-consortium.org/}{https://aquila-consortium.org/}.



\bibliographystyle{mnras}
\bibliography{research}



\newpage
\appendix

\section{Milky-Way r~band extinction distribution}
\label{app:ext}
The Milky-Way dust extinction has a non-negligible effect on observed magnitudes, as high as $\sim 1$ mag in the most extreme cases. To account for this one could use a Milky-Way dust map, apply a cut to the ZTF footprint and to the regions with $A_\mathrm{V} < 1$ mag to match the \sne data and obtain the resulting distribution of extinction in $r$-band $A_r$ over the sky. However this approach does not account for the difference in cadence over the sky or potential angular inhomogeneities in the SNe sample.

\begin{figure}
    \centering
    \includegraphics[width=\linewidth]{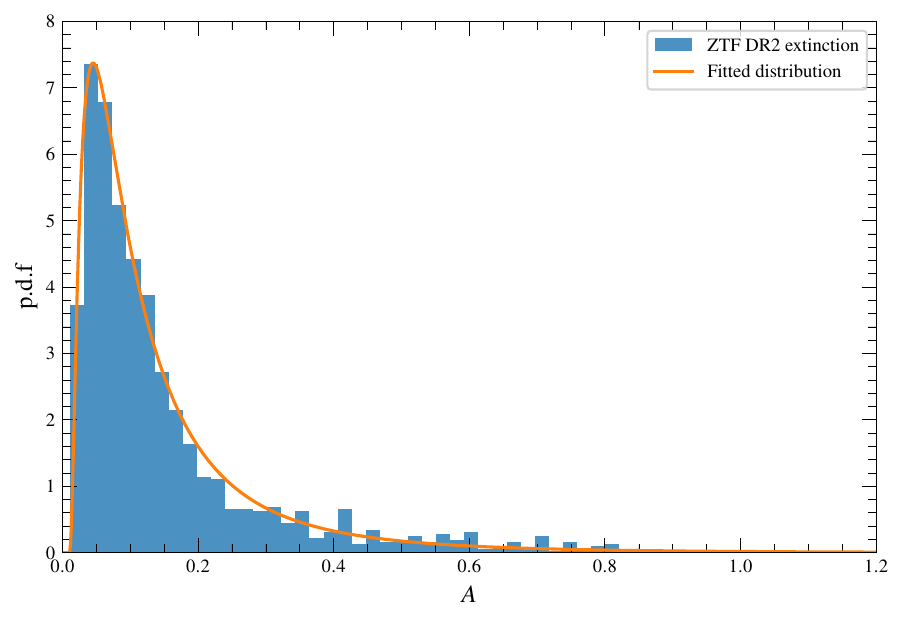}
    \caption{Model for the extinction distribution, the DR2 sample used in this analysis consists only of low extinction ($A_\mathrm{V} < 1$ mag), low redshift ($z < 0.06$) \sne.}
    \label{fig:extinction_distribution}
\end{figure}

To accommodate these two effects, we decided instead to directly fit the extinctions observed for the volume complete sample with $A_V<1.0$. The distribution of extinctions for this sample is represented in Fig. \ref{fig:extinction_distribution}.

We choose to model this distribution using a modified Log-normal distribution with the following p.d.f.
\begin{equation}
    h(A_r; a, b, \sigma) = f(\frac{A_r - a}{b}; 0, \sigma)
\end{equation}
where $f(x;\mu,\sigma)$ is the p.d.f of the usual Log-normal distribution
\begin{equation}
    f(x; \mu, \sigma) = \frac{1}{x\sigma\sqrt{2\pi}} e^{-\frac{1}{2}\qty(\frac{\ln(x) - \mu}{\sigma})^2}
\end{equation}
Using the \texttt{scipy} library, we fit this distribution to the data and obtain the following parameters
\begin{equation}
    \label{eq:fitext}
    a=0.0095 \pm 0.0008 \qc b=0.089 \pm 0.003 \qc\sigma=0.96 \pm 0.02
\end{equation}

\section{ZTF Cadence}
\label{app:cadence}
To account for coverage and cadence related effects, we made use of the fraction of the observed time covering an event location $f_{\mathrm{obstime},i}$ in our rate model (see Sec. \ref{sec:efficiency}).

To get the exact value of $f_{\mathrm{obstime},i}$, one would need to clone the observed transient with different times of maximum, simulate their lightcurves according to the observation logs and infer the fraction of those that would be detected. However we can greatly reduce the computation time by making use of the fact that the Northern Sky Survey is a 3-days cadence survey, and that a 3-days cadence cycle provides a high enough sampling to observe and characterize a transient. As such we group the observation logs on a 3-days cycle basis and considering a location to be observed if a field containing it was covered during this cycle. The fraction of the observing time covering the event location is thus
\begin{equation}
    f_{\mathrm{obstime},i} = \frac{N^\mathrm{cycles}_i}{N^\mathrm{cycles}_\mathrm{tot}}
\end{equation}
where $N^\mathrm{cycles}_i$ is the number of cycles where the event location was covered and $N^\mathrm{cycles}_\mathrm{tot}\simeq308$ is the total number of 3-days cycles during the DR2 operations.

\begin{figure}
    \centering
    \includegraphics[width=\linewidth]{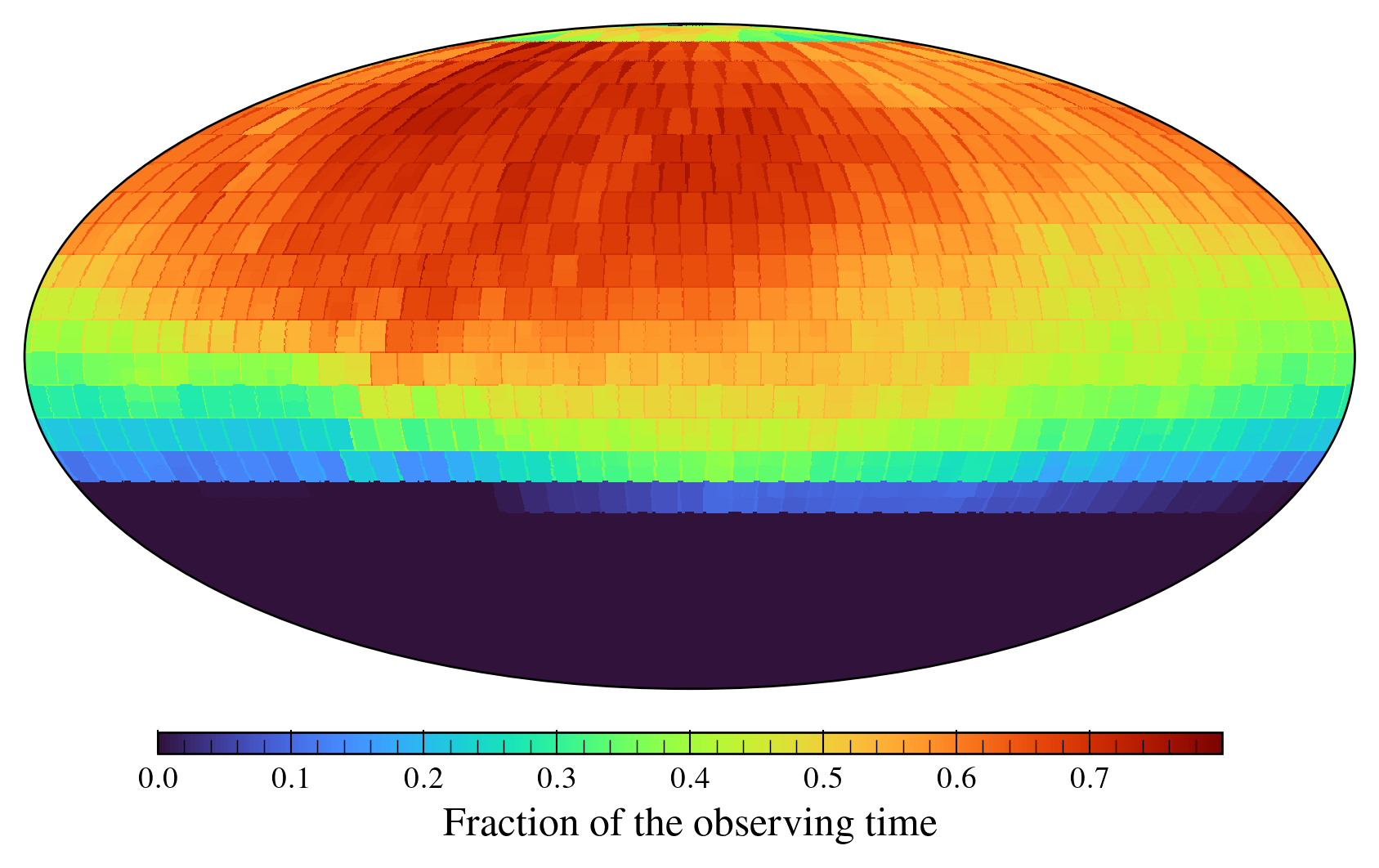}
    \caption{Weights used for the cadence}
    \label{fig:cadence}
\end{figure}

A map of the resulting fractions on the sky is shown in Fig. \ref{fig:cadence}. Regions with overlapping fields show the highest cadence, $\sim80\%$ of the total duration, while low declination locations were observed only $\sim10$--$20\%$ of the time due to seasonal variability. It is also worth noting the relatively low fraction of the northern polar region compared to nearby fields at only $\sim50$--$60\%$, mainly due to a change in operation as observation of fields with $\delta > 80\degr$ could only start in 2019. The mean active sky fraction we recover is $f_\mathrm{eff\ area} = 0.38$. It differs slightly from the value of $0.35$ used in \citet{perley2020}, which could be explained by the different durations considered and observation conditions.

On average, sky locations available to ZTF have been observed around $50\%$ of the total survey time. Comparing it to the observing time fraction of specific fields or regions leads to expect a maximum enhancement in the observed number of \snia of a factor $\sim 1.6$ for regions where several fields overlap or a factor $\sim 1.4$ for the individual fields that have been visited the most. The variation in cadence thus cannot alone explain the increase in the observed number of \sneia tied to specific structures, especially given that some of the increase seen in rates do not match the fields with higher cadence -- for instance Hercules a has been covered $~66\%$ of the survey time and Leo $~55\%$ while the rate in Hercules a is more than twice the one of Leo.

\section{Impact of Host redshift availability and redshift uncertainties on the rate}
\label{app:hostz}

As our analysis aims to estimate potential redshift dependencies of the \snia rate, it is crucial to have well-measured redshifts to avoid cross-bin contamination and obtain the correct weights through the survey efficiency. While around $62.7\%$ of the events have an associated host galaxy redshift with a very low uncertainty -- obtained either through host galaxy redshift measurement or emission lines in a non-SEDm spectra, some events have a redshift uncertainty close to the chosen bin size of $\Delta z =0.005$. Those redshifts were obtained either through emission lines in a SEDm spectra ($\delta z \sim 0.0015$) or via \snia template matching with \texttt{SNID} ($\delta z \sim 0.0036$). It is important to note that the fraction of \sne with a host redshift is itself redshift-dependent (see Fig. \ref{fig:app_zsource}).

To estimate potential biases arising from redshift uncertainties, several options are available. The first one is restricting the sample to the \sneia with available host-redshifts and add an additional correction to the efficiencies $\epsilon_i$ using the fraction of events having a host redshift measurement $f_\mathrm{host\ z}(z_i)$ at a given redshift. A second method is to keep using the full sample and reshuffle the redshifts according to their uncertainties, assuming their distribution to be Gaussian.

\begin{figure}
    \centering
    \includegraphics[width=\linewidth]{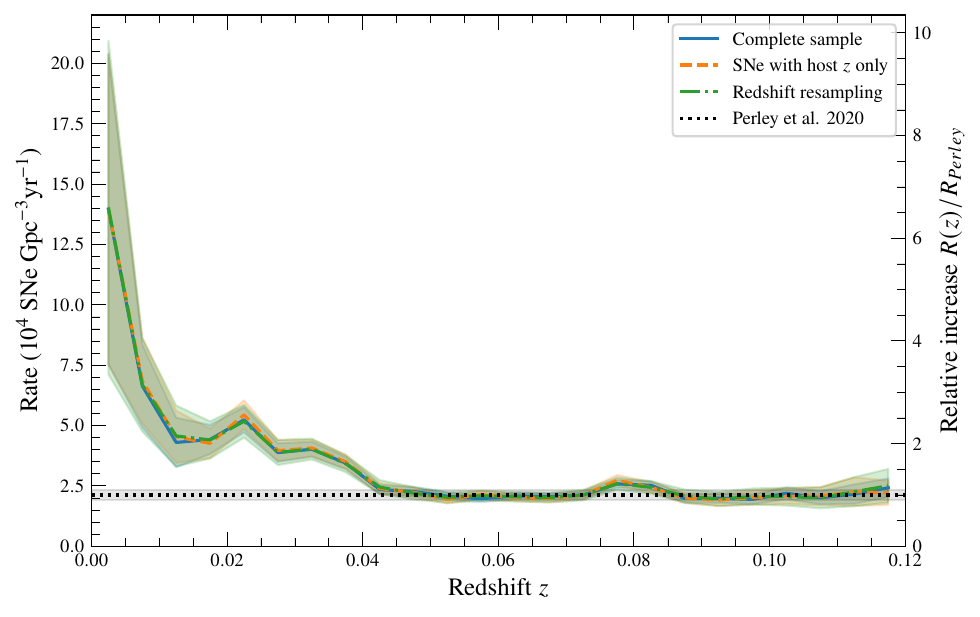}
    \caption{Reconstructed rate using either all the events, only those with a host redshift, or taking into account the redshift uncertainties in the efficiencies.}
    \label{fig:app_redshift_host}
\end{figure}

\begin{figure}
    \centering
    \includegraphics[width=\linewidth]{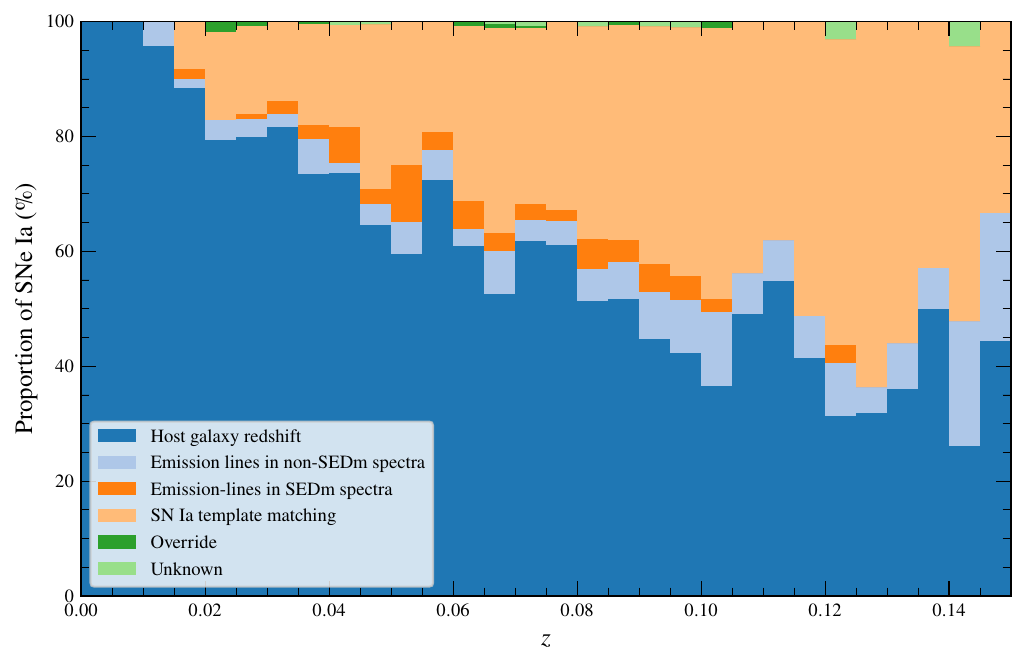}
    \caption{Distribution of the redshift source reported in the ZTF DR2 across the redshift bins considered. Host galaxy and non-SEDm emission lines have low uncertainties while SEDm emission lines and template matching have uncertainties close to the bin size}
    \label{fig:app_zsource}
\end{figure}

The redshift-dependent rates obtained from those two methods are shown in Fig. \ref{fig:app_redshift_host} and are well within the errorbars of the analysis using the full sample. This similarity is the mainly due to the sample being dominated by \snia with a host redshift measurement, especially at low redshifts. The \sneia without one can be seen as an additional noise upon this signal as their increased uncertainties shift them in or out of potential features.

\section{Weighting schemes for Manticore-based sampling}
\label{app:halo_weights}

When sampling the \sne positions from halos, different weighting scheme can be adopted. The simplest one is weighting all halos the same, \ie considering that the probability of a \snia occurring has no dependency with the halo properties. This scheme already reproduces the LSS formed by the halos and in particular overdense supercluster-like environments as they contain more sub-halos than less dense environments. A more realistic scheme consists in weighting each halo proportionally to their mass, as more massive halos are generally expected to host more massive galaxies and have a higher stellar mass.

The use of the \manticore halos catalogues also enables us to quantify the effect of cross-shell contamination coming from halos close to the redshift bins edges, whose infall motion onto a structure pulls to another shell. As infall velocities can reach $v\sim 1500$km/s it could lead to a displacement of the observed redshift by $\Delta z_\mathrm{infall} \sim 0.005$ compared to the cosmological one, similar to the bin size used throughout this article. This would cause an artificial enhancement of excesses tied to structures and potential contamination between shells. Another similar effect comes from the virial motion of the \sneia within their host halo. The typical velocity dispersion in a halo is of order $v \sim 1000$ km/s, leading to a displacement of the observed redshift by an additional $\Delta z_{vir} \sim 0.003$. However this last effect will smooth the redshift distribution and dilute the effect of well-localised structures, on average flattening fluctuations rather than producing new ones.

\begin{figure}
    \centering
    \includegraphics[width=\linewidth]{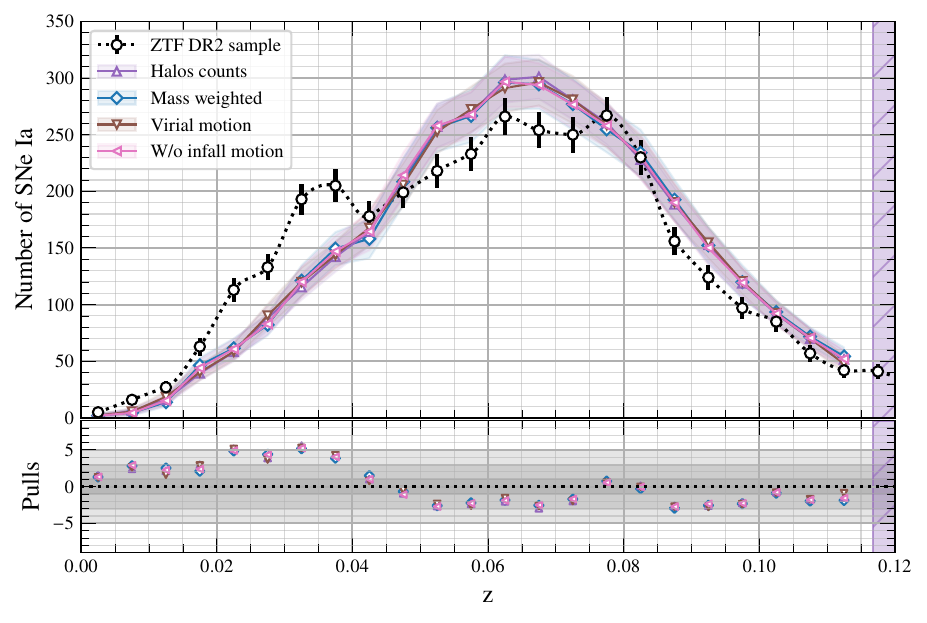}
    \caption{Rescaled redshift distributions for $N=100$ simulated samples for the \manticore-local halo catalogues i) without any weighting, ii) weighted by their mass iii) including virial motion with the mass weighting and iv) only considering the cosmological redshift (\ie neglecting any peculiar motion such as structure infall). The observed counts from the DR2 samples are shown as a reference.}
    \label{fig:app_halo_weights}
\end{figure}

A comparison of the redshift distributions of $N=100$ samples for those different schemes is shown in Fig. \ref{fig:app_halo_weights}, rescaled to match on average the total number of \sneia in the DR2. All produces very similar redshift distributions, deviating only slightly from the uniform sampling scheme. Their reduced chi-squares over the redshift range $z\leq 0.115$ is reported in Tab. \ref{tab:chi2_manticore}. The main improvement comes from the addition of the halos mass, while using only the halo counts or adding the virial motion to the mass weighting only marginally improves the fit. Although the scale of fluctuations does not allow to match the observed counts the mass-weighted sampling recovers the approximate redshift range of observed excesses (\eg $z\sim 0.015$--$0.025$, $z\sim0.03$--$0.04$ and even $z\sim0.08$). Cross shell contamination from infall velocities is negligible and well within the error bars at low redshift, although slight deviations can be observed between the simulations including infall velocities and the one that doesn't around $z\sim0.015$, $z\sim0.04$. With larger samples this effect could become a concern.

\begin{table}
    \centering
    \begin{tabular}{ccc}
         Sampling scheme & $\chi^2_\mathrm{d.o.f.}$ & Equivalent significance ($\sigma$)\\
         \hline
         Uniform & 7.68 & 10\\
         Halos counts & 7.38 & 9.7 \\
         Mass weighted & 6.64 & 9.1 \\
         Virial motion & 6.51 & 8.9 \\
         Without peculiar velocities & 6.97 & 9.4 \\
         \hline
    \end{tabular}
    \caption{Reduced chi-squares for the different sampling schemes over the redshift range $z\leq 0.115$}
    \label{tab:chi2_manticore}
\end{table}

\section{\snia angular distribution from \manticore}
\label{app:angular_manticore}
\begin{figure*}
    \centering
    \includegraphics[width=\linewidth]{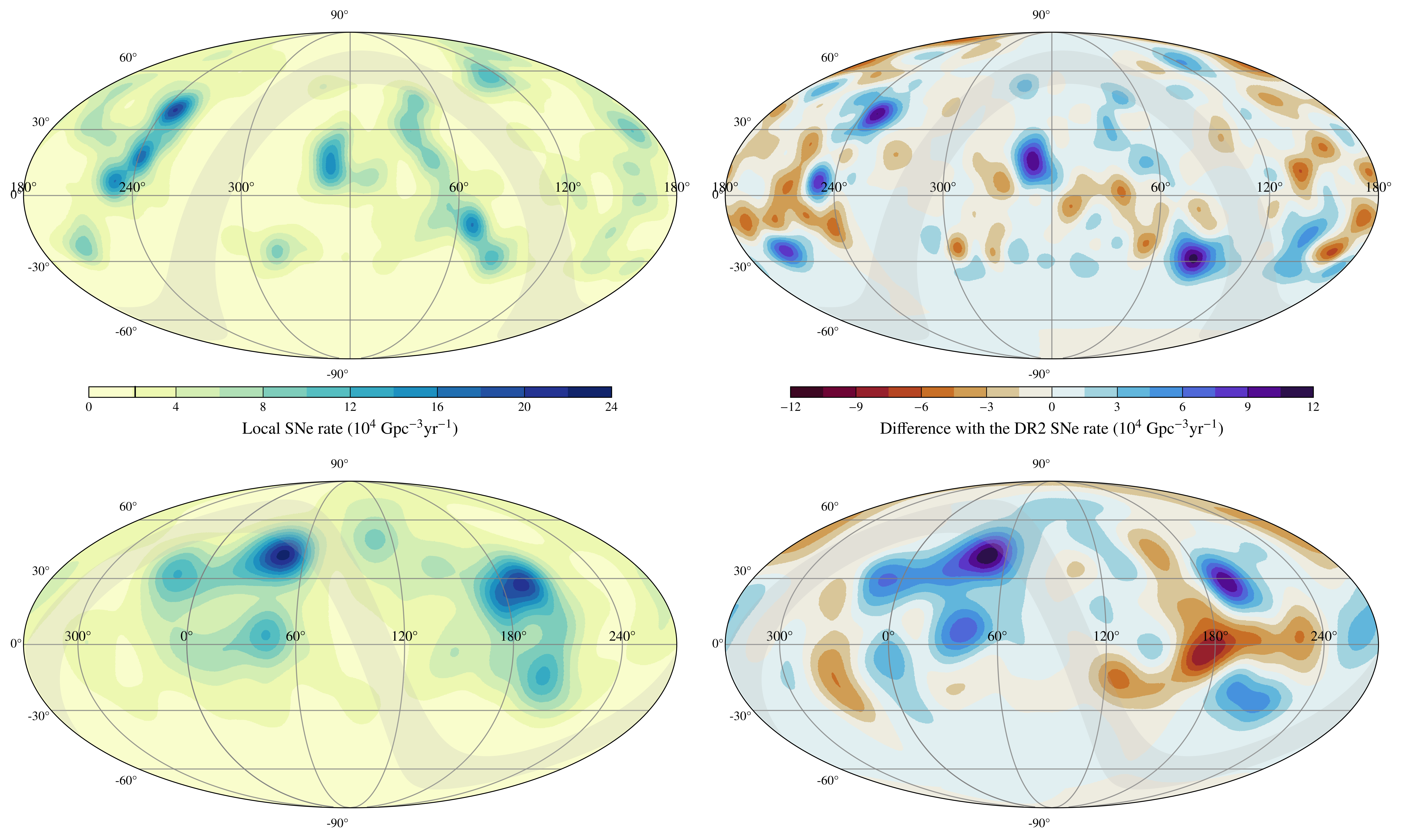}
    \caption{\textit{Left:} Mean reconstructed rate from \snia simulations using the \manticore-based sampling scheme, for the redshift slices $0.03\leq z\leq 0.04$ (\textit{upper}) and $0.015\leq z \leq 0.025$ (\textit{lower}). \textit{Right:} Corresponding deviation from the ZTF DR2 inferred maps (see Figs \ref{fig:0.015-0.025 slice} and \ref{fig:0.03-0.04 slice}).}
    \label{fig:manticore_skymaps}
\end{figure*}

While sampling \snia position based on \manticore halos allows to recover the angular distribution caused by the presence structures, this method does not reproduce the correct distribution. The angular distribution from such simulations using the redshift-dependent \snia rate inferred on the ZTF DR2 sample is shown in Fig. \ref{fig:manticore_skymaps}. While the main structures like Hercules, Perseus or Coma have indeed a higher rate, their scale and hierarchy do not match the observation from Figs. \ref{fig:0.015-0.025 slice} and \ref{fig:0.03-0.04 slice}. In these simulations the Perseus supercluster has a higher \snia rate than the Coma supercluster, Hercules a and Hercules b have similar rates and the Coma cluster (\abell{1656}) shows higher rates than the Leo Cluster (\abell{1367}). Therefore combining the redshift-dependent rate with the \manticore halo catalogues is not sufficient to fully describe the spatial distribution of \sneia.

\bsp	
\label{lastpage}
\end{document}